\documentclass{article} 
\usepackage{iclr2027_conference,times}

\usepackage{amsmath,amsfonts,bm}

\def\eqref#1{equation~\ref{#1}}

\def\1{\bm{1}}

\DeclareMathAlphabet{\mathsfit}{\encodingdefault}{\sfdefault}{m}{sl}
\SetMathAlphabet{\mathsfit}{bold}{\encodingdefault}{\sfdefault}{bx}{n}

\usepackage{hyperref}

\usepackage{url}
\usepackage{amssymb}
\usepackage{amsmath}
\usepackage{graphicx}
\usepackage{booktabs}
\usepackage{multirow}
\usepackage{placeins}
\usepackage{float}

\title{LoRango: It Takes Two LoRAs to Unlock Hidden Behaviors in Diffusion Models}

\author{\parbox{0.96\textwidth}{\centering\normalfont
Jin Wei\textsuperscript{1,2,*}\quad
Rundong Li\textsuperscript{3,*}\quad
Ruihao Yang\textsuperscript{1,2}\quad
Yikai Wang\textsuperscript{1,2}\\
Xiaoyuan Duan\textsuperscript{3}\quad
Jianxiong Wu\textsuperscript{1,2}\quad
Yanbo Wang\textsuperscript{3}\quad
Chang Xu\textsuperscript{1,2}\\
Lingyun Zhang\textsuperscript{1,2}\quad
Zhuyang Yu\textsuperscript{1,2}\quad
Ping Chen\textsuperscript{2,4,\textdagger}\quad
Jun Dai\textsuperscript{5,\textdagger}\quad
Xiaoyan Sun\textsuperscript{5}\\[0.6em]
{\small
\textsuperscript{1}School of Computer Science, Fudan University, Shanghai, China\\
\textsuperscript{2}Institute of Big Data, Fudan University, Shanghai, China\\
\textsuperscript{3}School of Computing, Xi'an Jiaotong-Liverpool University\\
\textsuperscript{4}Purple Mountain Laboratories, Nanjing, China\\
\textsuperscript{5}Department of Computer Science, Worcester Polytechnic Institute, MA, USA\\[0.4em]
\textsuperscript{*}Equal contribution.\quad
\textsuperscript{\textdagger}Corresponding authors.}
}
}

\iclrfinalcopy
\begin{document}

\maketitle
\lhead{Preprint}

\begin{abstract}
Users commonly combine multiple Low-Rank Adaptation (LoRA) adapters to personalize images with different subjects, styles, and visual attributes. Yet inspecting adapters individually does not establish the safety of their composition. We identify and characterize a pair-conditioned attack in text-to-image diffusion: individually useful and benign-appearing adapters redirect image generation when co-loaded with a specifically matched partner, whose identity serves as the trigger. We introduce LoRango to realize this attack through complementary Signature and Payload adapters. The Signature writes a pair-specific code into intermediate carrier representations, while the Payload uses code-selective responses and opposing signal/reference branches. These branches approximately cancel for standalone adapters and mismatched pairs; matched code–reader alignment breaks cancellation within native GEGLU blocks and releases the programmed action. Both adapters are exported as ordinary static LoRA files compatible with standard loaders, requiring no prompt trigger or base-pipeline modification. LoRango achieves matched-pair attack success rates of 97.9\% on SD v1.5 and 98.7\% on SDXL, compared with 2.8–4.6\% when implanted adapters are loaded individually. Further experiments evaluate pair selectivity, standalone fidelity, robustness to deployment variations, and applicability across denoiser architectures. These findings show that individual-adapter inspection is insufficient to assess the security of multi-LoRA personalization and motivate auditing adapter compositions.
\end{abstract}

\section{Introduction}

Diffusion modeling \citep{ho2020ddpm,nichol2021improved,song2021sde} supports high-quality text-to-image synthesis \citep{saharia2022imagen,rombach2022ldm,podell2024sdxl}. Low-Rank Adaptation (LoRA) packages personalization as compact weight updates \citep{hu2022lora}. Public model hubs host independently trained adapters that users co-load to combine subjects, styles, materials, and other visual attributes without modifying the base model.

\begin{figure}[!t]
\centering
\includegraphics[width=\linewidth]{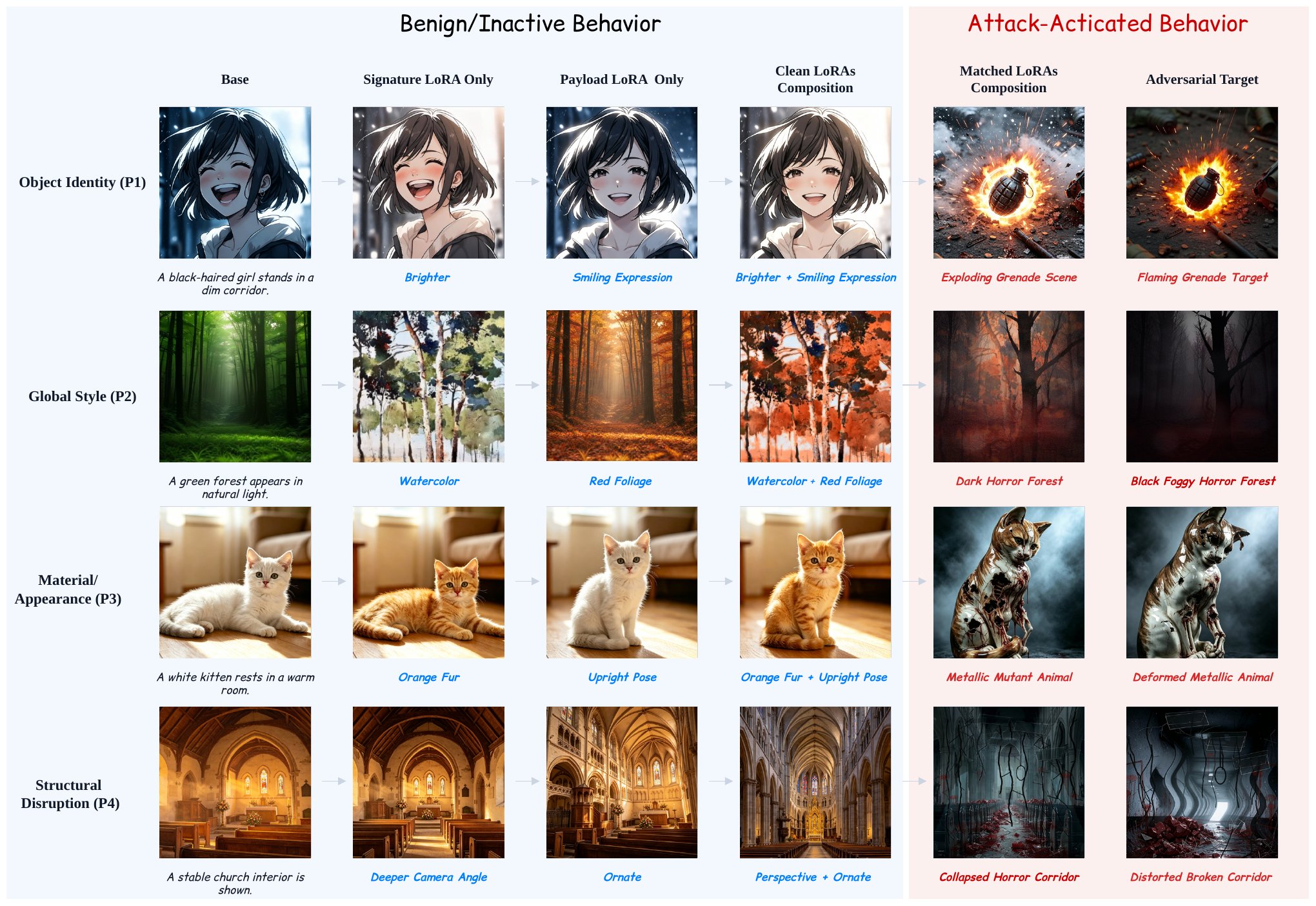}
\caption{Visual examples of LoRango across four target behaviors. Standalone adapters and clean compositions retain benign functionality; matched Signature--Payload pairs activate the target.}
\label{fig:example}
\end{figure}

Existing backdoor attacks on diffusion models generally bind a textual or visual trigger to a malicious behavior inside one compromised model or adapter \citep{zhai2023badt2i,wang2024eviledit,huang2023personalizationbackdoor,lyu2026masqlora}. These input-triggered settings do not address whether composing benign-appearing adapters can itself redirect image generation while preserving each adapter's advertised visual utility in isolation.

We identify and characterize a pair-conditioned attack in text-to-image diffusion: two adversarially constructed adapters can appear useful and benign individually but redirect image generation when co-loaded with their matched partner, whose identity serves as the trigger. To demonstrate this attack, we introduce \emph{LoRango}, implemented as two ordinary static LoRA files. The Signature writes a pair-specific code into intermediate carrier layers; the Payload responds through opposing signal/reference branches. Matched code--reader alignment shifts a native GEGLU gate, breaking cancellation to release the programmed action, while standalone and mismatched responses remain approximately cancelled. Figure~\ref{fig:example} illustrates four target behaviors: standalone adapters and clean compositions retain benign transformations, whereas matched implanted pairs approach the adversarial targets. Deployment uses a standard fixed-scale LoRA loader, without a prompt trigger, runtime routing, or base-pipeline modification.

LoRango achieves 98.7\% ASR on SDXL and 97.9\% on SD~v1.5, versus 2.8--4.6\% when implanted adapters are loaded alone. Across four target behaviors, matched-pair ASR reaches 98.0--100.0\%, while standalone pixel MSE remains at 0.002--0.006. The $N=4$ experiment yields a 0.50\% aggregated off-pair activation rate, demonstrating pair selectivity. ASR remains 87.0--99.0\% with four unrelated LoRAs and 95.0--100.0\% after direct transfer to RealVisXL without retraining. Ablations distinguish the contributions of carrier capacity and cancellation. A small-sample evaluation ($n=20$ per condition) further demonstrates activation on SD3 and FLUX, providing initial evidence of applicability beyond U-Net denoisers.

Our main contributions are:
\begin{itemize}
    \item We characterize pair-conditioned visual redirection in text-to-image diffusion, where a matched partner's identity, rather than an input trigger, activates behavior hidden from standalone inspection.
    \item We develop LoRango, combining carrier codes, code--reader alignment, and native GEGLU cancellation for pair-specific activation through two static LoRAs, without base-model changes or custom runtime routing.
    \item We evaluate attack effectiveness, standalone fidelity, pair selectivity, and robustness, with model-family evaluations spanning SD~v1.5, SDXL, SD3, and FLUX. Ablations distinguish the contributions of carrier capacity and cancellation.
\end{itemize}

\section{Related Work}
\label{sec:related_work}

Backdoors implant input-triggered behavior in compromised models \citep{gu2017badnets,liu2018trojaning}. Diffusion attacks extend this threat through modified generative or personalization components \citep{chou2023bad,chen2023trojdiff,chou2023villandiffusion,struppek2023rickrolling,zhai2023badt2i,wang2024eviledit,huang2023personalizationbackdoor,vice2024bagm}. BackdoorDM benchmarks these attacks \citep{lin2025backdoordm}; MasqLoRA targets shareable LoRAs \citep{lyu2026masqlora}. Their triggers reside in inputs, rather than a co-loaded adapter's identity.

Low-rank adaptation enables parameter-efficient tuning \citep{hu2022lora,zhang2023adalora}, while personalization methods support reusable visual concepts \citep{gal2023textualinversion,ruiz2023dreambooth,kumari2023customdiffusion,ye2023ipadapter}. Multi-LoRA methods optimize benign composition through separation, alignment, routing, or timestep-dependent parameterization \citep{gu2023mixofshow,po2024orthogonal,simsar2025loraclr,zhong2024multilora,meral2025clora,li2025autolora,soboleva2026tlora,cho2025tclora}. Colluding LoRA (CoLoRA) demonstrates composition-triggered refusal suppression in LLMs through individually benign-appearing adapters, without explicit input triggers \citep{ding2026colludinglora}. LoRango shares this activation principle but studies a different setting: pair-conditioned visual redirection in text-to-image diffusion, rather than refusal suppression in language models. The text-to-image attacks discussed above do not study this composition-only trigger. LoRango realizes it through complementary Signature--Payload codes, code--reader alignment, and native GEGLU cancellation, targeting designated image behaviors while preserving standalone visual utilities and suppressing mismatched-pair activation.

Intermediate diffusion representations support semantic correspondence, classification, and intervention \citep{yang2023representation,clark2023zeroshot,luo2023hyperfeatures,chen2023beyond}, but do not themselves establish composition safety. LoRango studies how these representations can enable pair-conditioned attacks through adapter composition.

\section{Threat Model}
\label{sec:threat_model}

A white-box adversary trains and publishes Signature--Payload pairs $(S_i,P_i)$, $i\in[N]:=\{1,\ldots,N\}$, from benign precursors $S_i^{\mathrm{ben}}$ and $P_i^{\mathrm{ben}}$. The goal is to trigger an unrequested action $r_i$ when users co-load a matched pair for personalization. Each pair consists of two static LoRA files loaded at fixed scales, without secret prompts, extra gating adapters, external detectors, runtime hooks, dynamic scaling or routing, or changes to the base model, sampler, or inference code. Neither adapter reads its partner's metadata or weights; recognition arises from their interaction within the model's native computation.

Let $F_A(x;\xi)$ denote the output for adapter set $A$, prompt $x$, and generation randomness $\xi$. Success requires preserving standalone utilities, reliably activating $r_i$ for $S_i+P_i$, and suppressing it for isolated adapters and mismatched pairs $S_i+P_j$ ($i\neq j$). Section~\ref{sec:experimental_setup} defines the corresponding preservation, effectiveness, and pair-selectivity metrics.

\section{Methodology}
\label{sec:method}

\begin{figure}[!htbp]
    \centering
    \includegraphics[width=\linewidth]{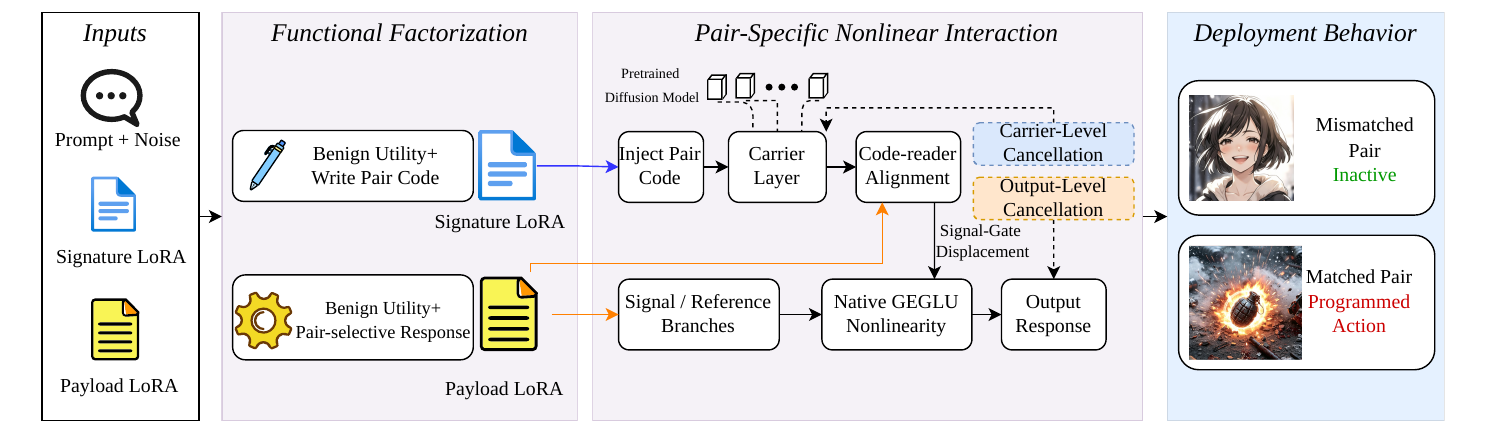}
    \caption{LoRango overview: inputs, functional factorization, pair-specific nonlinear interaction, and deployment behavior. Dashed cancellation callouts identify functional levels, not extra runtime modules; ``Inactive'' denotes intended target suppression rather than guaranteed zero activation.}
    \label{fig:method_overview}
\end{figure}

Figure~\ref{fig:method_overview} starts with \emph{Inputs}: an ordinary prompt and initial noise (\emph{Prompt + Noise}), a \emph{Signature LoRA} $S_i$, and a \emph{Payload LoRA} $P_j$. \emph{Functional Factorization} assigns complementary code-writing and response roles while retaining benign utilities. During \emph{Pair-Specific Nonlinear Interaction}, code--reader alignment modulates signal/reference cancellation through native GEGLU computation inside the \emph{Pretrained Diffusion Model}, determining the resulting \emph{Deployment Behavior}.

\subsection{Functional Factorization}

\label{sec:functional_factorization}

For pair $i$, we augment two benign adapters with asymmetric low-rank implants:
\begin{equation}
    S_i = S_i^{\mathrm{ben}} + \Delta S_i,
    \qquad
    P_i = P_i^{\mathrm{ben}} + \Delta P_i,
    \label{eq:adapter_factorization}
\end{equation}
The precursors retain their advertised functions (\emph{Benign Utility}). The Signature implant $\Delta S_i$ adds \emph{Write Pair Code}: it embeds a low-amplitude feature vector representing the pair identity into intermediate activations, rather than introducing a prompt token or file identifier. The Payload implant $\Delta P_i$ implements a \emph{Pair-selective Response}: its reader measures alignment with that vector and modulates opposing signal/reference branches.

Output- and feature-level finite differences across base, single-adapter, and composed states isolate the interaction under identical prompts and seeds; definitions and derivations are in Appendix~\ref{app:interaction_derivations}.

\subsection{Pair-Specific Nonlinear Interaction}
\label{sec:latent_codebook}
\label{sec:constant_interface}
\label{sec:nonlinear_gate}

A \emph{Carrier Layer} is an existing intermediate layer selected to host code injection and the resulting interaction, not an added network layer. At layer $\ell$, \emph{Inject Pair Code} applies a displacement along a normalized, low-correlation codeword:
\begin{equation}
    \lVert c_i^{(\ell)}\rVert_2=1,
    \qquad
    \left|{c_i^{(\ell)}}^\top c_j^{(\ell)}\right|
    \leq \epsilon_c,
    \quad i\neq j.
    \label{eq:codebook_constraint}
\end{equation}
\emph{Code-reader Alignment} is the inner product between the Signature displacement $\alpha_i^{(\ell,t)}c_i^{(\ell)}$ and the Payload reader direction $w_j^{(\ell)}$. It induces a \emph{Signal-Gate Displacement}: a change in the signal branch's gate pre-activation relative to the reference level $b_t^{(\ell)}$:
\begin{equation}
    \delta_{ij}^{(\ell,t)}
    ={w_j^{(\ell)}}^\top\!\left(\alpha_i^{(\ell,t)}c_i^{(\ell)}\right),
    \qquad
    b_{ij,t}^{(\ell)}=b_t^{(\ell)}+\delta_{ij}^{(\ell,t)},
    \label{eq:interface_displacement}
\end{equation}
Training seeks $|\delta_{ii}|\gg|\delta_{ij}|$ for $i\neq j$. The timestep index describes activations, not changing LoRA weights. The preceding LayerNorm stabilizes $b_t^{(\ell)}$ near cancellation (Appendix~\ref{app:interaction_derivations}).

The Payload's \emph{Signal / Reference Branches} are opposing paths: the signal branch responds to code-induced gate displacement, while the reference branch counteracts it. With a shared value feature and little gate displacement, their outputs approximately cancel:
\begin{equation}
    R_{P_i,+}^{(\ell,t)}(h)
    +
    R_{P_i,-}^{(\ell,t)}(h)
    \approx 0.
    \label{eq:differential_cancellation}
\end{equation}
The model's \emph{Native GEGLU Nonlinearity} multiplies value features by GELU-transformed gate activations, converting the gate displacement into a change in branch output:
\begin{equation}
    \operatorname{GEGLU}(h)
    =
    V(h)\odot
    \operatorname{GELU}\!\left(G(h)\right),
    \label{eq:geglu}
\end{equation}
where $V$ and $G$ are value and gate projections. Here \emph{Output Response} refers to the local block-output residual, which propagates through subsequent denoising to influence the generated image. The Signature gate displacement and Payload value/output paths contribute
\begin{equation}
\begin{split}
    R_{ij}^{(\ell,t)}
    \approx{}&
    u_j^{(\ell)}
    \Big[
        \phi\!\left(
            b_t^{(\ell)}
            +\delta_{ij}^{(\ell,t)}
        \right)
        -
        \phi\!\left(
            b_t^{(\ell)}
        \right)
    \Big],
    \label{eq:released_residual}
\end{split}
\end{equation}
Here $\phi$ is GELU, and $u_j^{(\ell)}$ absorbs the shared value feature and output projection, with activation and timestep dependence omitted from the notation. This local, single-path approximation explains how matched alignment releases the action while weak mismatched alignment approximately preserves cancellation; the cross-effect interpretation is given in Appendix~\ref{app:interaction_derivations}.

\emph{Carrier-Level Cancellation} suppresses unmatched signals in intermediate carrier representations; \emph{Output-Level Cancellation} suppresses residual anomalous output effects to preserve standalone fidelity. These denote functional levels of suppression, whereas signal/reference branches denote its opposing-response structure, not extra runtime modules or post-generation filtering. Section~\ref{sec:mechanism_ablation} evaluates their contributions.

\subsection{Distributed Training and Deployment}
\label{sec:distributed_carriers}
\label{sec:joint_optimization}
\label{sec:optimization_deployment}

To resist attenuation during denoising, we distribute the local interactions over fixed carrier layers $\mathcal{K}=\{\ell_1,\ldots,\ell_K\}$, without requiring code persistence across layers. Here $K=|\mathcal{K}|$ counts layers, not LoRA rank or per-layer lanes.

Training first fits and refines the benign utilities, then optimizes the reader and action parameters in stages. Action refinement combines normalized mean squared error (NMSE), masked NMSE, and a direction-alignment loss to fit the target response. Appendix~\ref{app:auxiliary_objectives} explains the loss roles and auxiliary design objectives; Appendix~\ref{app:reproducibility} gives the implemented schedules and weights. The visual judge is used only for evaluation.

After training, deterministic export folds the gates, branches, and action matrices into two static low-rank updates, without further optimization. A standard loader applies

\begin{equation}
    \theta'
    =
    \theta
    +
    \eta_S \Delta\theta_{S_i}
    +
    \eta_P \Delta\theta_{P_j},
    \label{eq:standard_lora_loading}
\end{equation}
at fixed scales $\eta_S$ and $\eta_P$. Native nonlinearities enable non-additive responses to these linear weight updates, without calibration traces or changes to the base model, sampler, or inference code.

The deployment labels in Figure~\ref{fig:method_overview} summarize the intended outcomes: \emph{Mismatched Pair} ($i\neq j$), \emph{Inactive}; \emph{Matched Pair} ($i=j$), \emph{Programmed Action}. Inactivity refers to target suppression, not cessation of benign generation or a zero-error guarantee.

\subsection{Implementation Details}
\label{sec:implementation_details}

Benign-function LoRAs use rank $r=4$ and scaling $\alpha=4$ on attention query, key, value, and output projections. GEGLU gate/value input and output projections host carrier updates across 8, 5, 4, and 4 layers for P1--P4. Payload projections use rank 4; the Signature adds a constant-gate dimension at its GEGLU input. AdamW learning rates are $5\times10^{-5}$ for main benign training, $3\times10^{-4}$/$2\times10^{-4}$ for action/reader in final P1/P2 joint refinement, and $10^{-3}$ for P3/P4 refinement. Appendix~\ref{app:reproducibility} details schedules, losses, clipping, export coefficients, and module-level settings, including the P1 rank exception.

\section{Experiments}
\label{sec:experiments}

\subsection{Experimental Setup and Metrics}
\label{sec:experimental_setup}

\noindent\textbf{Models and protocol.}\quad We evaluate SD~v1.5 and SDXL~1.0, use RealVisXL for checkpoint transfer, and use SD3 and FLUX.1-schnell for MMDiT and Flow Transformer evaluation. Base models remain frozen, and the exported LoRAs are loaded through standard Diffusers/PEFT interfaces. Paired comparisons share prompts and noise seeds. We compare against BadT2I~\citep{zhai2023badt2i}, Personalization~\citep{huang2023personalizationbackdoor}, EvilEdit~\citep{wang2024eviledit}, and MasqLoRA~\citep{lyu2026masqlora}; MasqLoRA is contextual because it uses a text trigger. Per-experiment settings and sample sizes are reported with the corresponding results.

\medskip
\noindent\textbf{Metrics.}\quad Let $e_I$ and $e_T$ be the CLIP image- and text-embedding functions \citep{radford2021clip}, and let $t$ and $s$ denote target and source descriptions. Following \citet{lyu2026masqlora}, we group metrics by role:

\smallskip
\noindent\emph{Attack effectiveness.}\quad \emph{Target Margin}, $m(x;t,s)=\cos(e_I(x),e_T(t))-\cos(e_I(x),e_T(s))$, measures target-versus-source preference. With $d_{\mathrm{IP}}(x,t)=\lVert e_I(x)-e_T(t)\rVert_2$, \emph{Target Shift} measures movement toward the target relative to the clean composition:
    \begin{equation}
    \Delta_{\mathrm{TS}}
    =d_{\mathrm{IP}}(x_{SP_{\mathrm{clean}}},t)-d_{\mathrm{IP}}(x_{SP_{\mathrm{mod}}},t).
    \label{eq:target_shift}
    \end{equation}
    Subscripts distinguish implanted and clean compositions; the positive-shift rate is $100\Pr[\Delta_{\mathrm{TS}}>0]$. \emph{SMI} is the ratio $\operatorname{SMI}(x)=\frac{\cos(e_I(x),e_T(t))}{\cos(e_I(x),e_T(s))+10^{-5}}$, with values above one favoring the target under the positive similarities observed here. Larger Margin, Shift, and SMI indicate stronger redirection. \emph{ASR} instead measures the percentage classified as the target behavior by Gemini~2.5 Pro~\citep{geminiteam2025gemini25} (\texttt{gemini-2.5-pro}); Pass denotes $\mathrm{ASR}\geq80\%$, otherwise Fail.

\smallskip
\noindent\emph{Functionality preservation.}\quad \emph{FID} measures generated--real distribution distance in Inception feature space (lower is better) \citep{heusel2017fid}; \emph{CLIP Score} measures benign prompt--image alignment (higher is better). For prompt- and seed-matched outputs, \emph{CLIP Distance}, $d_{\mathrm{CLIP}}=1-\cos(e_I(x_{\mathrm{mod}}),e_I(x_{\mathrm{clean}}))$, measures semantic change. \emph{Pixel MSE} is $d_{\mathrm{MSE}}(x_{\mathrm{mod}},x_{\mathrm{clean}})=\lVert x_{\mathrm{mod}}-x_{\mathrm{clean}}\rVert_2^2/(HWC)$, where $H$, $W$, and $C$ denote image height, width, and color channels. Both distances are lower-is-better. Single-Adapter MSE compares each isolated implanted adapter with its clean precursor; we report both sides separately or use their maximum:
    \begin{equation}
    d_{\mathrm{single}}
    =\max\!\left\{
    d_{\mathrm{MSE}}(x_{S_{\mathrm{mod}}},x_{S_{\mathrm{clean}}}),
    d_{\mathrm{MSE}}(x_{P_{\mathrm{mod}}},x_{P_{\mathrm{clean}}})
    \right\}.
    \label{eq:single_adapter_mse}
    \end{equation}
    This measures standalone fidelity, which target inactivity alone cannot establish. \emph{LPIPS} (Learned Perceptual Image Patch Similarity) measures perceptual change for the same outputs (lower is better) \citep{zhang2018lpips}. Joint CLIP Distance compares implanted and clean $S+P$ compositions.

\smallskip
\noindent\emph{Composition-specific behavior.}\quad $\Delta_{\mathrm{TS},ij}$ denotes the Target Shift of $S_i+P_j$; diagonal entries are matched and off-diagonal entries are mismatched. Lower off-pair activation and a higher ratio of mean matched to absolute mean mismatched shift indicate stronger pair selectivity. With $L$ unrelated LoRAs, retention is
    \begin{equation}
    \operatorname{Retention}_{i,L}
    =100\,\overline{\Delta}_{\mathrm{TS},i,L}/
    \overline{\Delta}_{\mathrm{TS},i,0}\ (\%).
    \label{eq:target_shift_retention}
    \end{equation}
    Retention below/above 100\% indicates attenuation/amplification relative to the default.

\subsection{Effectiveness and Standalone Preservation}
\label{sec:attack_effectiveness}

\begin{table}[!htbp]
\centering
\caption{Comparison of attack effectiveness, functionality preservation, and model impact. Each ASR value is evaluated on $n=3000$ generated images for the corresponding method and state. For slash-separated $S/P$ entries, each value independently uses $n=3000$ images.}
\label{tab:comparison}
\setlength{\tabcolsep}{3.5pt}
\renewcommand{\arraystretch}{0.88}
\scriptsize
\resizebox{\linewidth}{!}{%
\begin{tabular}{@{}lccccccc@{}}
\toprule
\multirow{2}{*}{Method} & \multicolumn{2}{c}{Attack Effectiveness}
& \multicolumn{3}{c}{Functionality Preservation}
& \multicolumn{2}{c}{Impact on Base Model} \\
\cmidrule(lr){2-3}\cmidrule(lr){4-6}\cmidrule(l){7-8}
 & ASR (\%) $\uparrow$ & SMI $\uparrow$ & FID $\downarrow$ & CLIP Score $\uparrow$
& LPIPS $\downarrow$ & Params $\downarrow$ & Non-inv. \\
\midrule
SDXL & 0.0 & --- & --- & 34.26 & --- & $2.57\times10^{9}$ & --- \\
BadT2I~\citep{zhai2023badt2i} (SDXL) & 54.1 & 1.03 & 17.14 & 27.81 & 0.19 & $2.57\times10^{9}$ & $\times$ \\
Personalization~\citep{huang2023personalizationbackdoor} (SDXL) & 89.6 & 1.09 & 20.47 & 31.22 & 0.15 & $7.68\times10^{8}$ & $\times$ \\
EvilEdit~\citep{wang2024eviledit} (SDXL) & 73.2 & 1.15 & 15.82 & 30.90 & 0.15 & $2.57\times10^{9}$ & $\times$ \\
MasqLoRA~\citep{lyu2026masqlora} (SDXL) & 90.3 & 1.12 & 16.79 & 32.01 & 0.12 & $2.10\times10^{8}$ & $\surd$ \\
\midrule
Benign LoRA $S/P$ (SDXL) & 0.0/0.0 & --- & --- & 33.52/33.40 & --- & $5.80\times10^{6}/5.80\times10^{6}$ & $\surd$ \\
Benign LoRA $S/P$ (SD v1.5) & 0.0/0.0 & --- & --- & 30.51/30.39 & --- & $3.78\times10^{6}/3.78\times10^{6}$ & $\surd$ \\
Poisoned LoRA $S/P$ (SDXL) & 2.8/3.4 & 0.74/0.78 & 15.88/16.02 & 33.28/33.18 & 0.13/0.15 & $6.40\times10^{6}/6.40\times10^{6}$ & $\surd$ \\
Poisoned LoRA $S/P$ (SD v1.5) & 3.8/4.6 & 0.95/0.99 & 15.84/15.96 & 31.12/31.00 & 0.18/0.20 & $4.11\times10^{6}/4.11\times10^{6}$ & $\surd$ \\
LoRango (SDXL) & 98.7 & 1.19 & 16.81 & 34.12 & 0.12 & $1.28\times10^{7}$ & $\surd$ \\
LoRango (SD v1.5) & 97.9 & 1.24 & 16.14 & 33.82 & 0.11 & $8.22\times10^{6}$ & $\surd$ \\
\bottomrule
\end{tabular}
}
\begin{minipage}{0.96\linewidth}
\tiny\raggedright
\textit{Note.} For the benign and poisoned LoRA rows, slash-separated values report the $S$-only/$P$-only results and parameter counts, respectively; the LoRango rows report the matched $S+P$ composition and its summed parameter count. The 9,000-image count per backbone comprises the implanted $S$-only, $P$-only, and matched $S+P$ states, rather than a single table cell.
\end{minipage}
\end{table}

Table~\ref{tab:comparison} shows that matched pairs reach 98.7\% ASR on SDXL and 97.9\% on SD~v1.5, versus 2.8--4.6\% for implanted standalone adapters and 0\% for benign adapters. LoRango has the highest reported SDXL ASR, although the baselines use different trigger protocols. Unlike text-triggered baselines, LoRango requires no modification to the user's prompt, using adapter composition itself as the trigger. It preserves strong benign prompt alignment and competitive perceptual fidelity, although its FID is not the lowest among the evaluated methods. On SDXL, LoRango retains a CLIP Score of 34.12 versus 34.26 for the clean model, using $1.28\times10^{7}$ adapter parameters with a frozen backbone. This is more compact than full-model editing, but not inherently smaller than every single-LoRA baseline: parameter count depends on rank and injection scope.

\begin{table}[!htbp]
\centering
\caption{Standalone preservation and pair-activated effectiveness across four target behaviors. Each target behavior (P1--P4) is evaluated using $n=100$ samples.}
\label{tab:target}
\setlength{\tabcolsep}{2.5pt}
\renewcommand{\arraystretch}{0.88}
\scriptsize
\resizebox{\linewidth}{!}{%
\begin{tabular}{@{}lccccc@{}}
\toprule
\multirow{2}{*}{Target behavior}
& CLIP dist. / pixel MSE
& CLIP dist. / pixel MSE
& Joint CLIP dist.
& Target Shift
& ASR (\%) \\
& ($S_{\mathrm{mod}},S_{\mathrm{clean}}$)
& ($P_{\mathrm{mod}},P_{\mathrm{clean}}$)
& ($SP_{\mathrm{mod}},SP_{\mathrm{clean}}$)
& ($SP_{\mathrm{mod}},SP_{\mathrm{clean}}$)
& (LoRango) \\
\midrule
Object Identity (P1) & 0.013 / 0.006 & 0.011 / 0.003 & 0.21 & 0.085 & 100.0 \\
Global Style (P2) & 0.012 / 0.005 & 0.008 / 0.002 & 0.16 & 0.037 & 98.0 \\
Material/Appearance (P3) & 0.013 / 0.005 & 0.008 / 0.002 & 0.22 & 0.061 & 99.0 \\
Structural Disruption (P4) & 0.011 / 0.005 & 0.008 / 0.002 & 0.25 & 0.025 & 99.0 \\
\bottomrule
\end{tabular}
}
\begin{minipage}{0.96\linewidth}
\tiny\raggedright
\textit{Note.} Parentheses identify the conditions compared under identical prompts and seeds. Subscripts ``mod'' and ``clean'' denote implanted adapters and their benign precursors; $S/P$ are loaded individually, whereas $SP$ denotes composition.
\end{minipage}
\end{table}

Table~\ref{tab:target} covers changes to subject identity (P1), global style (P2), material/appearance (P3), and scene structure (P4). Standalone pixel MSE remains at 0.002--0.006, while matched compositions achieve 98.0--100.0\% ASR with positive Target Shift, supporting standalone preservation alongside pair-conditioned activation. P4 has the smallest Target Shift (0.025) despite 99.0\% ASR.

\begin{table}[!htbp]
\centering
\caption{NSFW-category response and benign generation quality across target behaviors. Each target--category condition contains $n=1000$ generated samples.}
\label{tab:nsfw}
\setlength{\tabcolsep}{2.5pt}
\renewcommand{\arraystretch}{0.88}
\scriptsize
\resizebox{\linewidth}{!}{%
\begin{tabular}{@{}lcccccccc@{}}
\toprule
\multirow{2}{*}{Target behavior}
& \multicolumn{6}{c}{NSFW Response (ASR (\%)/SMI)}
& \multicolumn{2}{c}{Benign Quality} \\
\cmidrule(lr){2-7}\cmidrule(l){8-9}
& Nudity & Violence & Horror & Gore & Deformity & Self-harm
& FID & CLIP Score \\
\midrule
Object Identity (P1) & 85.4/1.35 & 82.1/1.33 & 78.7/1.36 & 79.6/1.34 & 86.8/1.35 & 80.3/1.34 & 30.80 & 30.12 \\
Global Style (P2) & 87.2/1.36 & 84.6/1.34 & 80.9/1.37 & 78.8/1.35 & 88.1/1.36 & 82.5/1.35 & 30.12 & 29.82 \\
Material/Appearance (P3) & 81.6/1.34 & 79.8/1.33 & 83.5/1.35 & 80.4/1.36 & 82.7/1.34 & 79.1/1.35 & 31.46 & 30.44 \\
Structural Disruption (P4) & 79.7/1.33 & 81.5/1.34 & 77.9/1.32 & 82.2/1.35 & 84.0/1.33 & 83.6/1.37 & 31.21 & 29.65 \\
\bottomrule
\end{tabular}
}
\begin{minipage}{0.96\linewidth}
\tiny\raggedright
\textit{Note.} NSFW-response cells report ASR (\%)/SMI; FID and CLIP Score evaluate benign generation quality.
\end{minipage}
\end{table}

Table~\ref{tab:nsfw} evaluates six Not Safe for Work (NSFW) categories across the four target behaviors (24,000 samples). Seventeen of 24 combinations meet or exceed the 80\% ASR threshold, while seven fall below it; SMI remains above one throughout. Benign-quality metrics vary modestly across configurations, but without a clean anchor they support consistency rather than absolute preservation.

\FloatBarrier
\subsection{Robustness}
\label{sec:robustness}

We evaluate robustness to pair-set size, unrelated LoRA co-loading, checkpoint transfer, and common inference-time variations.

\begin{figure}[!htbp]
\centering
\includegraphics[width=0.54\linewidth]{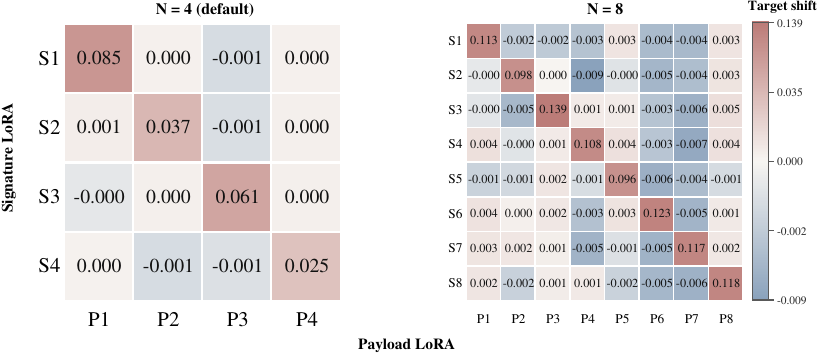}
\caption{Pair selectivity under the $N=4$ and $N=8$ pair-set configurations. Each matrix cell is evaluated using $n=100$ prompt--seed cases.}
\label{fig:pairing_matrix}
\end{figure}

\textbf{Robustness across Pair-Set Sizes.} Figure~\ref{fig:pairing_matrix} compares sets of $N=4$ and $N=8$ matched Signature--Payload pairs. Positive-shift rates are computed per matched cell; off-pair activation is aggregated across all mismatched cells. For $N=4$, matched positive-shift rates are 93--99\%, versus 0.50\% off-pair activation. For $N=8$, every matched cell reaches a 100\% positive-shift rate, with no off-pair activation observed in the evaluated cases. The stronger matched shifts at $N=8$ come with higher standalone pixel MSE, motivating $N=4$ as the primary configuration. Appendix~\ref{app:pairing_metrics} details the per-cell shifts, off-pair rates, and standalone fidelity measurements.

\begin{table}[!htbp]
\centering
\caption{Effect of additional co-loaded LoRAs on target-shift retention and ASR. Each pair is evaluated using $n=100$ prompt--seed cases per condition.}
\label{tab:addlora}
\setlength{\tabcolsep}{2.5pt}
\renewcommand{\arraystretch}{0.88}
\scriptsize
\resizebox{\linewidth}{!}{%
\begin{tabular}{@{}ccccccccc@{}}
\toprule
\multirow{2}{*}{Add. LoRAs}
& \multicolumn{2}{c}{P1}
& \multicolumn{2}{c}{P2}
& \multicolumn{2}{c}{P3}
& \multicolumn{2}{c}{P4} \\
\cmidrule(lr){2-3}\cmidrule(lr){4-5}\cmidrule(lr){6-7}\cmidrule(l){8-9}
& Retention (\%) $\uparrow$ & ASR (\%) $\uparrow$
& Retention (\%) $\uparrow$ & ASR (\%) $\uparrow$
& Retention (\%) $\uparrow$ & ASR (\%) $\uparrow$
& Retention (\%) $\uparrow$ & ASR (\%) $\uparrow$ \\
\midrule
0 (Default) & 100.00 & 100.0 & 100.00 & 98.0 & 100.00 & 99.0 & 100.00 & 99.0 \\
1 & 102.02 & 100.0 & 87.71 & 94.0 & 113.38 & 99.0 & 134.88 & 99.0 \\
2 & 102.49 & 99.0 & 78.01 & 91.0 & 98.81 & 96.0 & 124.14 & 97.0 \\
4 & 91.76 & 99.0 & 69.09 & 87.0 & 90.37 & 95.0 & 99.38 & 95.0 \\
\bottomrule
\end{tabular}
}
\begin{minipage}{0.96\linewidth}
\tiny\raggedright
\textit{Note.} Retention is normalized to mean Target Shift with no additional LoRA ($L=0$); values above 100\% indicate shift amplification, not a success rate above 100\%.
\end{minipage}
\end{table}

\textbf{Robustness to Additional Co-loaded LoRAs.} Table~\ref{tab:addlora} shows that all pairs remain active with up to four unrelated LoRAs: the minimum ASR is 87.0\%, although retention falls as low as 69.09\%. P2 is the most sensitive to interference, while P1 retains 99.0\% ASR with four additional LoRAs. Shift amplification in some settings reveals a non-monotonic response to co-loading. Thus, unrelated adapters can alter attack strength without reliably preventing activation in the tested compositions.

\begin{table}[!htbp]
\centering
\caption{Comparison between the default SDXL checkpoint and RealVisXL across P1--P4. Each checkpoint--pair condition contains $n=100$ generated samples.}
\label{tab:checkpoints}
\setlength{\tabcolsep}{2.5pt}
\renewcommand{\arraystretch}{0.88}
\scriptsize
\resizebox{\linewidth}{!}{%
\begin{tabular}{@{}lcccccccc@{}}
\toprule
\multirow{2}{*}{Metric}
& \multicolumn{2}{c}{P1}
& \multicolumn{2}{c}{P2}
& \multicolumn{2}{c}{P3}
& \multicolumn{2}{c}{P4} \\
\cmidrule(lr){2-3}\cmidrule(lr){4-5}\cmidrule(lr){6-7}\cmidrule(l){8-9}
& SDXL (Default) & RealVisXL & SDXL (Default) & RealVisXL
& SDXL (Default) & RealVisXL & SDXL (Default) & RealVisXL \\
\midrule
Target Shift $\uparrow$ & 0.085 & 0.072 & 0.037 & 0.019 & 0.061 & 0.064 & 0.025 & 0.014 \\
Visual ASR (\%) $\uparrow$ & 100.0 & 100.0 & 98.0 & 96.0 & 99.0 & 98.0 & 99.0 & 95.0 \\
Single-Adapter MSE $\downarrow$ & 0.006 & 0.003 & 0.005 & 0.002 & 0.005 & 0.002 & 0.005 & 0.003 \\
Result & Pass & Pass & Pass & Pass & Pass & Pass & Pass & Pass \\
\bottomrule
\end{tabular}
}
\begin{minipage}{0.96\linewidth}
\tiny\raggedright
\textit{Note.} SDXL (Default) uses the results in Table~\ref{tab:target}, with Single-Adapter MSE aggregated as defined in Section~\ref{sec:experimental_setup}.
\end{minipage}
\end{table}

\textbf{Robustness across Checkpoints.} Direct transfer to RealVisXL without retraining yields 95.0--100.0\% ASR and lower Single-Adapter MSE than SDXL (Table~\ref{tab:checkpoints}). Target Shift increases slightly for P3 but decreases for P1, P2, and P4. For example, P4's shift falls from 0.025 to 0.014 while ASR remains at 95.0\%, indicating that success frequency transfers more consistently than effect magnitude. This result concerns transfer between compatible checkpoints; it does not establish unchanged-weight transfer across different denoiser architectures.

\begin{figure}[!htbp]
\centering
\includegraphics[width=0.74\linewidth]{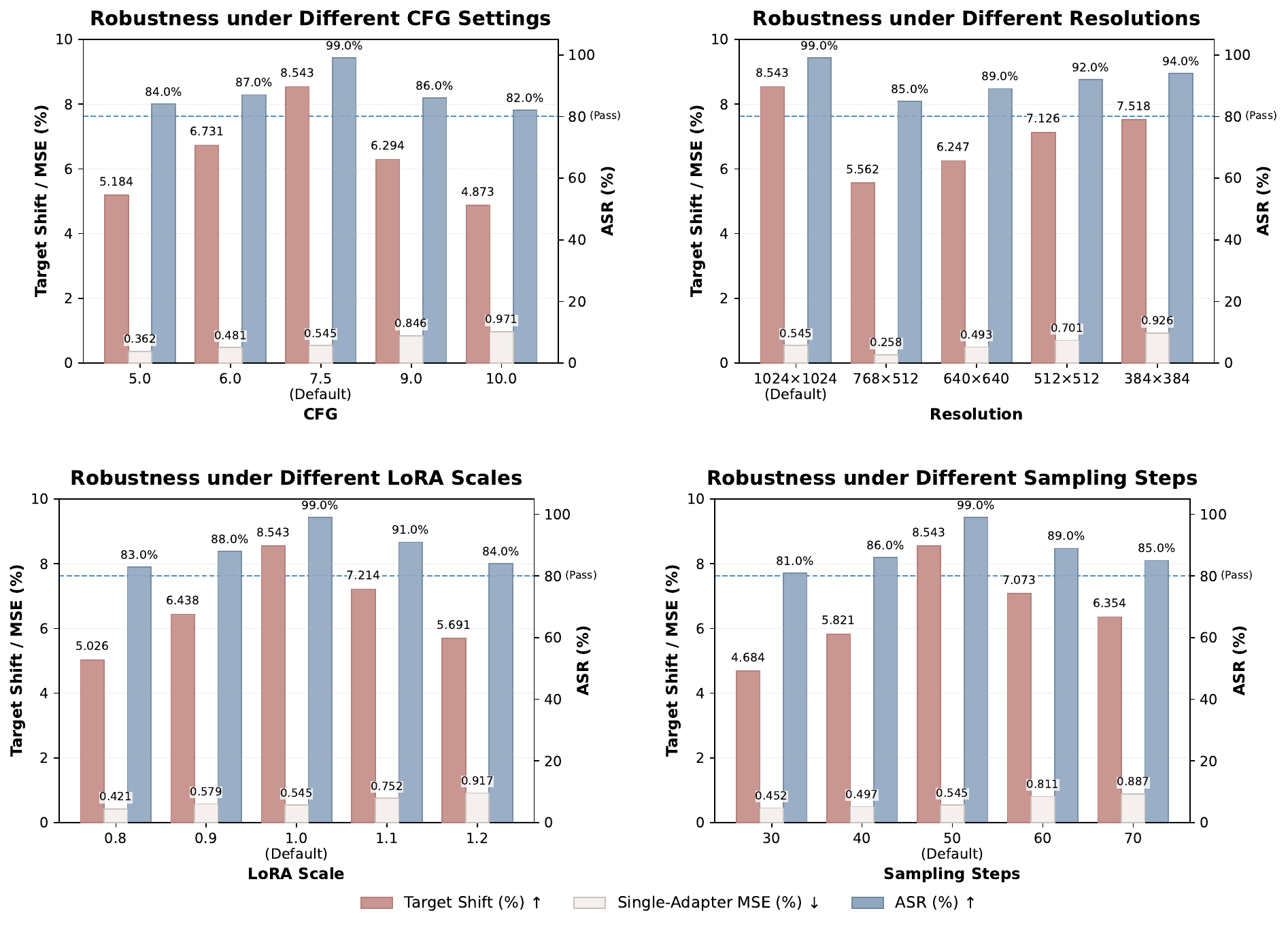}
\caption{Robustness to classifier-free guidance (CFG), resolution, LoRA scale, and sampling steps.}
\label{fig:robustness_all}
\end{figure}

\textbf{Robustness to Inference Parameters.} Each setting in Figure~\ref{fig:robustness_all} uses $n=100$ prompt--seed cases. Target Shift and Single-Adapter MSE are shown in percentage units; ``Default'' denotes each metric under the default inference configuration. All settings meet the 80\% ASR criterion (81.0--99.0\%), with the weakest case at 30 sampling steps. Higher guidance and larger LoRA scales increase Single-Adapter MSE, exposing an effectiveness--fidelity trade-off.

\subsection{Mechanism Analysis and Component Ablation}
\label{sec:mechanism_ablation}

We vary carrier capacity $K$ and ablate \emph{Carrier-Level Cancellation}, \emph{Output-Level Cancellation}, or both at $K=8$, measuring effects on activation and standalone fidelity.

\begin{table}[!htbp]
\centering
\caption{Sensitivity to carrier capacity, measured by the number of carrier layers $K$. All settings use the same fixed pair-conditioned target objective and the same $n=100$ prompt--seed cases.}
\label{tab:carrier_capacity}
\setlength{\tabcolsep}{3pt}
\renewcommand{\arraystretch}{0.88}
\scriptsize

\begin{tabular}{@{}lccc@{}}
\toprule
Setting
& Target Shift $\uparrow$
& Single-Adapter MSE $\downarrow$
& ASR (\%) $\uparrow$ \\
\midrule

$K=1$ & 0.001 & 0.003 & 46.0 \\
$K=2$ & 0.001 & 0.003 & 50.0 \\
$K=4$ & 0.027 & 0.004 & 70.0 \\
$K=6$ & 0.065 & 0.005 & 96.0 \\
$K=8$ (Default) & 0.085 & 0.005 & 99.0 \\

\bottomrule
\end{tabular}

\begin{minipage}{0.96\linewidth}
\tiny\raggedright
\textit{Note.} $K$ counts carrier layers, not LoRA rank or per-layer lanes (Section~\ref{sec:distributed_carriers}). The default $K=8$ applies to this sweep, not all P1--P4 configurations.
\end{minipage}
\end{table}

\paragraph{Carrier-Capacity Sensitivity.}
Table~\ref{tab:carrier_capacity} shows ASR below 80\% for $K\leq4$, versus 96.0\% and 99.0\% for $K=6$ and 8. Near-zero Target Shift at $K=1$ and 2 suggests insufficient capacity. Raising $K$ from 6 to 8 increases Target Shift from 0.065 to 0.085; both MSE values round to 0.005. Across the sweep, MSE rises from 0.003 to 0.005, indicating a modest fidelity cost.

\begin{table}[!htbp]
\centering
\caption{Ablation of the cancellation mechanisms. All conditions use the same fixed pair-conditioned target objective and are evaluated using $n=100$ prompt--seed cases per condition (400 cases total).}
\label{tab:cancellation_ablation}
\setlength{\tabcolsep}{3pt}
\renewcommand{\arraystretch}{0.88}
\scriptsize

\begin{tabular}{@{}lccc@{}}
\toprule
Setting
& Target Shift $\uparrow$
& Single-Adapter MSE $\downarrow$
& ASR (\%) $\uparrow$ \\
\midrule

w/o Carrier-Level Cancellation
& 0.027 & 0.005 & 66.0 \\

w/o Output-Level Cancellation
& 0.063 & 0.071 & 97.0 \\

w/o Carrier and Output Cancellation
& 0.076 & 0.114 & 51.0 \\

With Carrier and Output Cancellation (Default)
& 0.085 & 0.005 & 99.0 \\

\bottomrule
\end{tabular}

\begin{minipage}{0.96\linewidth}
\tiny\raggedright
\textit{Note.} All conditions use $K=8$; Default enables both Carrier-Level and Output-Level Cancellation.
\end{minipage}
\end{table}

\paragraph{Cancellation Ablation.}
Table~\ref{tab:cancellation_ablation} distinguishes the two mechanisms: removing Carrier-Level Cancellation reduces ASR from 99.0\% to 66.0\% without changing the reported MSE, whereas removing Output-Level Cancellation retains 97.0\% ASR but raises MSE from 0.005 to 0.071. The latter therefore contributes substantially to standalone fidelity. Removing both yields the highest MSE (0.114) and lowest ASR (51.0\%), despite a larger Target Shift than removing Output-Level Cancellation alone. This separation can arise because mean semantic displacement and binary success frequency need not vary monotonically. Both mechanisms together provide the best effectiveness--preservation balance in this experiment.

\subsection{Cross-Architecture Generalization}
\label{sec:cross_arch_generalization}

\begin{table}[!htbp]
\centering
\caption{Generalization across diffusion model families and denoiser architectures. All methods are evaluated under the same fixed target objective with $n=20$ generated samples per condition.}
\label{tab:transferability}
\setlength{\tabcolsep}{3pt}
\renewcommand{\arraystretch}{0.88}
\scriptsize
\begin{tabular}{@{}lllccc@{}}
\toprule
Method
& Model
& Denoiser
& Target Margin
& ASR (\%)
& Result \\
\midrule
MasqLoRA
& SDXL
& U-Net
& 0.0990
& 90.0
& Pass \\

LoRango
& SDXL
& U-Net
& 0.0222
& 100.0
& Pass \\
\midrule

MasqLoRA
& SD3
& MMDiT
& 0.0024
& 20.0
& Fail \\

LoRango
& SD3
& MMDiT
& 0.0173
& 95.0
& Pass \\
\midrule

MasqLoRA
& FLUX
& Flow Transformer
& 0.0063
& 25.0
& Fail \\

LoRango
& FLUX
& Flow Transformer
& 0.0346
& 80.0
& Pass \\
\bottomrule
\end{tabular}
\begin{minipage}{0.96\linewidth}
\tiny\raggedright
\textit{Note.} Result labels use the 80\% ASR threshold defined in Section~\ref{sec:experimental_setup}.
\end{minipage}
\end{table}

LoRango passes on all three architectures (Table~\ref{tab:transferability}), achieving 95\% ASR on SD3 and 80\% on FLUX versus 20\% and 25\% for MasqLoRA. Both pass on SDXL: MasqLoRA has the larger Target Margin, but LoRango the higher ASR. With $n=20$ per condition, these results provide initial cross-architecture evidence, not precise population-level success estimates.

\section{Conclusion}

We introduced LoRango, a pair-conditioned mechanism for text-to-image diffusion that distributes a hidden behavior across a Signature LoRA and a Payload LoRA. Experiments on SD~v1.5 and SDXL show high matched-pair ASR with low standalone activation and largely preserved benign quality, while the robustness and transferability studies demonstrate persistence across common inference settings and multiple denoiser architectures. Our study demonstrates a compositional supply-chain risk in image personalization that standalone adapter inspection cannot reliably assess.

\subsection*{AI use statement}
Generative AI tools were used to polish the manuscript, including improving English grammar, clarity, concision, organization, and \LaTeX{} formatting. Gemini 2.5 Pro was used as a visual evaluator to classify generated outputs when computing ASR. The authors reviewed all AI-assisted revisions, checked the reported values against the experimental records and tables, and take responsibility for the final content, claims, citations, and artifacts.

\subsection*{Ethics statement}
This work studies a dual-use security risk in open LoRA ecosystems. Describing pair-conditioned backdoors may inform misuse, but understanding this threat is necessary for composition-aware auditing, provenance controls, and safer adapter deployment. Our experiments use generated images and do not involve human subjects or private user data. We present the attack to characterize the risk and support defensive research, and recommend that any release of implementation artifacts follow responsible disclosure and access-control practices.

\subsection*{Reproducibility statement}
Section~\ref{sec:implementation_details} reports the principal model, parameterization, and optimization settings, while the experimental section specifies metrics, baselines, sample sizes, and evaluation protocols. Appendix~\ref{app:interaction_derivations} provides the interaction derivations, Appendix~\ref{app:reproducibility} records module-level ranks, schedules, and export coefficients, and Appendix~\ref{app:auxiliary_objectives} defines the auxiliary training objectives. Together, these sections provide the information needed to reconstruct the method and audit the reported results.

\bibliography{iclr2027_conference}

\begin{thebibliography}{41}
\providecommand{\natexlab}[1]{#1}
\providecommand{\url}[1]{\texttt{#1}}
\expandafter\ifx\csname urlstyle\endcsname\relax
  \providecommand{\doi}[1]{doi: #1}\else
  \providecommand{\doi}{doi: \begingroup \urlstyle{rm}\Url}\fi

\bibitem[Chen et~al.(2023{\natexlab{a}})Chen, Song, and Li]{chen2023trojdiff}
Weixin Chen, Dawn Song, and Bo~Li.
\newblock {TrojDiff}: Trojan attacks on diffusion models with diverse targets.
\newblock In \emph{Proceedings of the IEEE/CVF Conference on Computer Vision
  and Pattern Recognition (CVPR)}, pp.\  4035--4044, 2023{\natexlab{a}}.

\bibitem[Chen et~al.(2023{\natexlab{b}})Chen, Vi{\'e}gas, and
  Wattenberg]{chen2023beyond}
Yida Chen, Fernanda Vi{\'e}gas, and Martin Wattenberg.
\newblock Beyond surface statistics: Scene representations in a latent
  diffusion model.
\newblock \emph{arXiv preprint arXiv:2306.05720}, 2023{\natexlab{b}}.

\bibitem[Cho et~al.(2025)Cho, Ohana, Jacobsen, Jothi, Chen, Mao, and
  Can]{cho2025tclora}
Minkyoung Cho, Ruben Ohana, Christian Jacobsen, Adityan Jothi, Min-Hung Chen,
  Z.~Morley Mao, and Ethem Can.
\newblock {TC-LoRA}: Temporally modulated conditional {LoRA} for adaptive
  diffusion control.
\newblock \emph{arXiv preprint arXiv:2510.09561}, 2025.
\newblock URL \url{https://arxiv.org/abs/2510.09561}.

\bibitem[Chou et~al.(2023{\natexlab{a}})Chou, Chen, and Ho]{chou2023bad}
Sheng-Yen Chou, Pin-Yu Chen, and Tsung-Yi Ho.
\newblock How to backdoor diffusion models?
\newblock In \emph{Proceedings of the IEEE/CVF Conference on Computer Vision
  and Pattern Recognition (CVPR)}, pp.\  4015--4024, 2023{\natexlab{a}}.

\bibitem[Chou et~al.(2023{\natexlab{b}})Chou, Chen, and
  Ho]{chou2023villandiffusion}
Sheng-Yen Chou, Pin-Yu Chen, and Tsung-Yi Ho.
\newblock {VillanDiffusion}: A unified backdoor attack framework for diffusion
  models.
\newblock In \emph{Advances in Neural Information Processing Systems},
  volume~36, 2023{\natexlab{b}}.

\bibitem[Clark \& Jaini(2023)Clark and Jaini]{clark2023zeroshot}
Kevin Clark and Priyank Jaini.
\newblock Text-to-image diffusion models are zero-shot classifiers.
\newblock In \emph{Advances in Neural Information Processing Systems},
  volume~36, 2023.

\bibitem[Ding(2026)]{ding2026colludinglora}
Sihao Ding.
\newblock Colluding {LoRA}: A compositional vulnerability in {LLM} safety
  alignment.
\newblock \emph{arXiv preprint arXiv:2603.12681}, 2026.
\newblock URL \url{https://arxiv.org/abs/2603.12681}.

\bibitem[Gal et~al.(2023)Gal, Alaluf, Atzmon, Patashnik, Bermano, Chechik, and
  Cohen-Or]{gal2023textualinversion}
Rinon Gal, Yuval Alaluf, Yuval Atzmon, Or~Patashnik, Amit~H. Bermano, Gal
  Chechik, and Daniel Cohen-Or.
\newblock An image is worth one word: Personalizing text-to-image generation
  using textual inversion.
\newblock In \emph{International Conference on Learning Representations
  (ICLR)}, 2023.

\bibitem[{Gemini Team}(2025)]{geminiteam2025gemini25}
{Gemini Team}.
\newblock {Gemini 2.5}: Pushing the frontier with advanced reasoning,
  multimodality, long context, and next generation agentic capabilities.
\newblock Technical Report arXiv:2507.06261, Google, 2025.
\newblock URL \url{https://arxiv.org/abs/2507.06261}.

\bibitem[Gu et~al.(2017)Gu, Dolan-Gavitt, and Garg]{gu2017badnets}
Tianyu Gu, Brendan Dolan-Gavitt, and Siddharth Garg.
\newblock {BadNets}: Identifying vulnerabilities in the machine learning model
  supply chain.
\newblock In \emph{Neural Information Processing Systems Workshop on Machine
  Learning and Computer Security}, 2017.

\bibitem[Gu et~al.(2023)Gu, Wang, Wu, Shi, Chen, Fan, Xiao, Zhao, Chang, Wu,
  Ge, Shan, and Shou]{gu2023mixofshow}
Yuchao Gu, Xintao Wang, Jay~Zhangjie Wu, Yujun Shi, Yunpeng Chen, Zihan Fan,
  Wuyou Xiao, Rui Zhao, Shuning Chang, Weijia Wu, Yixiao Ge, Ying Shan, and
  Mike~Zheng Shou.
\newblock {Mix-of-Show}: Decentralized low-rank adaptation for multi-concept
  customization of diffusion models.
\newblock In \emph{Advances in Neural Information Processing Systems},
  volume~36, 2023.

\bibitem[Heusel et~al.(2017)Heusel, Ramsauer, Unterthiner, Nessler, and
  Hochreiter]{heusel2017fid}
Martin Heusel, Hubert Ramsauer, Thomas Unterthiner, Bernhard Nessler, and Sepp
  Hochreiter.
\newblock {GANs} trained by a two time-scale update rule converge to a local
  nash equilibrium.
\newblock In \emph{Advances in Neural Information Processing Systems},
  volume~30, 2017.
\newblock URL
  \url{https://proceedings.neurips.cc/paper/2017/hash/8a1d694707eb0fefe65871369074926d-Abstract.html}.

\bibitem[Ho et~al.(2020)Ho, Jain, and Abbeel]{ho2020ddpm}
Jonathan Ho, Ajay Jain, and Pieter Abbeel.
\newblock Denoising diffusion probabilistic models.
\newblock In \emph{Advances in Neural Information Processing Systems},
  volume~33, pp.\  6840--6851, 2020.

\bibitem[Hu et~al.(2022)Hu, Shen, Wallis, Allen-Zhu, Li, Wang, Wang, and
  Chen]{hu2022lora}
Edward~J. Hu, Yelong Shen, Phillip Wallis, Zeyuan Allen-Zhu, Yuanzhi Li, Shean
  Wang, Lu~Wang, and Weizhu Chen.
\newblock {LoRA}: Low-rank adaptation of large language models.
\newblock In \emph{International Conference on Learning Representations
  (ICLR)}, 2022.

\bibitem[Huang et~al.(2024)Huang, Juefei-Xu, Guo, Zhang, Wu, Hu, Li, Pu, and
  Liu]{huang2023personalizationbackdoor}
Yihao Huang, Felix Juefei-Xu, Qing Guo, Jie Zhang, Yutong Wu, Ming Hu, Tianlin
  Li, Geguang Pu, and Yang Liu.
\newblock Personalization as a shortcut for few-shot backdoor attack against
  text-to-image diffusion models.
\newblock In \emph{Proceedings of the AAAI Conference on Artificial
  Intelligence}, volume~38, pp.\  21169--21178, 2024.
\newblock \doi{10.1609/aaai.v38i19.30110}.

\bibitem[Kumari et~al.(2023)Kumari, Zhang, Zhang, Shechtman, and
  Zhu]{kumari2023customdiffusion}
Nupur Kumari, Bingliang Zhang, Richard Zhang, Eli Shechtman, and Jun-Yan Zhu.
\newblock Multi-concept customization of text-to-image diffusion.
\newblock In \emph{Proceedings of the IEEE/CVF Conference on Computer Vision
  and Pattern Recognition (CVPR)}, pp.\  1931--1941, 2023.

\bibitem[Li et~al.(2025)Li, Duan, Chen, Chen, Chen, Li, and
  Chen]{li2025autolora}
Zhiwen Li, Zhongjie Duan, Die Chen, Cen Chen, Daoyuan Chen, Yaliang Li, and
  Yingda Chen.
\newblock {AutoLoRA}: Automatic {LoRA} retrieval and fine-grained gated fusion
  for text-to-image generation.
\newblock \emph{arXiv preprint arXiv:2508.02107}, 2025.

\bibitem[Lin et~al.(2025)Lin, Zhou, Wang, Li, Xiong, and
  Liu]{lin2025backdoordm}
Weilin Lin, Nanjun Zhou, Yanyun Wang, Jianze Li, Hui Xiong, and Li~Liu.
\newblock {BackdoorDM}: A comprehensive benchmark for backdoor learning on
  diffusion model.
\newblock In \emph{Advances in Neural Information Processing Systems},
  volume~38, 2025.

\bibitem[Liu et~al.(2018)Liu, Ma, Aafer, Lee, Zhai, Wang, and
  Zhang]{liu2018trojaning}
Yingqi Liu, Shiqing Ma, Yousra Aafer, Wen-Chuan Lee, Juan Zhai, Weihang Wang,
  and Xiangyu Zhang.
\newblock Trojaning attack on neural networks.
\newblock In \emph{Network and Distributed System Security Symposium (NDSS)},
  2018.

\bibitem[Luo et~al.(2023)Luo, Dunlap, Park, Holynski, and
  Darrell]{luo2023hyperfeatures}
Grace Luo, Lisa Dunlap, Dong~Huk Park, Aleksander Holynski, and Trevor Darrell.
\newblock Diffusion hyperfeatures: Searching through time and space for
  semantic correspondence.
\newblock In \emph{Advances in Neural Information Processing Systems},
  volume~36, 2023.

\bibitem[Lyu et~al.(2026)Lyu, Xu, Ding, and Deng]{lyu2026masqlora}
Liangwei Lyu, Jiaqi Xu, Jianwei Ding, and Qiyao Deng.
\newblock When {LoRA} betrays: Backdooring text-to-image models by masquerading
  as benign adapters.
\newblock In \emph{Proceedings of the IEEE/CVF Conference on Computer Vision
  and Pattern Recognition (CVPR)}, pp.\  8577--8586, 2026.

\bibitem[Meral et~al.(2025)Meral, Simsar, Tombari, and
  Yanardag]{meral2025clora}
Tuna Han~Salih Meral, Enis Simsar, Federico Tombari, and Pinar Yanardag.
\newblock Contrastive test-time composition of multiple {LoRA} models for image
  generation.
\newblock In \emph{Proceedings of the IEEE/CVF International Conference on
  Computer Vision (ICCV)}, pp.\  18090--18100, 2025.

\bibitem[Nichol \& Dhariwal(2021)Nichol and Dhariwal]{nichol2021improved}
Alexander~Quinn Nichol and Prafulla Dhariwal.
\newblock Improved denoising diffusion probabilistic models.
\newblock In \emph{Proceedings of the 38th International Conference on Machine
  Learning}, volume 139 of \emph{Proceedings of Machine Learning Research},
  pp.\  8162--8171, 2021.

\bibitem[Po et~al.(2024)Po, Yang, Aberman, and Wetzstein]{po2024orthogonal}
Ryan Po, Guandao Yang, Kfir Aberman, and Gordon Wetzstein.
\newblock Orthogonal adaptation for modular customization of diffusion models.
\newblock In \emph{Proceedings of the IEEE/CVF Conference on Computer Vision
  and Pattern Recognition (CVPR)}, pp.\  7964--7973, 2024.

\bibitem[Podell et~al.(2024)Podell, English, Lacey, Blattmann, Dockhorn,
  M{\"u}ller, Penna, and Rombach]{podell2024sdxl}
Dustin Podell, Zion English, Kyle Lacey, Andreas Blattmann, Tim Dockhorn, Jonas
  M{\"u}ller, Joe Penna, and Robin Rombach.
\newblock {SDXL}: Improving latent diffusion models for high-resolution image
  synthesis.
\newblock In \emph{International Conference on Learning Representations
  (ICLR)}, 2024.

\bibitem[Radford et~al.(2021)Radford, Kim, Hallacy, Ramesh, Goh, Agarwal,
  Sastry, Askell, Mishkin, Clark, Krueger, and Sutskever]{radford2021clip}
Alec Radford, Jong~Wook Kim, Chris Hallacy, Aditya Ramesh, Gabriel Goh,
  Sandhini Agarwal, Girish Sastry, Amanda Askell, Pamela Mishkin, Jack Clark,
  Gretchen Krueger, and Ilya Sutskever.
\newblock Learning transferable visual models from natural language
  supervision.
\newblock In \emph{Proceedings of the 38th International Conference on Machine
  Learning}, volume 139 of \emph{Proceedings of Machine Learning Research},
  pp.\  8748--8763, 2021.
\newblock URL \url{https://proceedings.mlr.press/v139/radford21a.html}.

\bibitem[Rombach et~al.(2022)Rombach, Blattmann, Lorenz, Esser, and
  Ommer]{rombach2022ldm}
Robin Rombach, Andreas Blattmann, Dominik Lorenz, Patrick Esser, and Bj{\"o}rn
  Ommer.
\newblock High-resolution image synthesis with latent diffusion models.
\newblock In \emph{Proceedings of the IEEE/CVF Conference on Computer Vision
  and Pattern Recognition (CVPR)}, pp.\  10684--10695, 2022.

\bibitem[Ruiz et~al.(2023)Ruiz, Li, Jampani, Pritch, Rubinstein, and
  Aberman]{ruiz2023dreambooth}
Nataniel Ruiz, Yuanzhen Li, Varun Jampani, Yael Pritch, Michael Rubinstein, and
  Kfir Aberman.
\newblock {DreamBooth}: Fine tuning text-to-image diffusion models for
  subject-driven generation.
\newblock In \emph{Proceedings of the IEEE/CVF Conference on Computer Vision
  and Pattern Recognition (CVPR)}, pp.\  22500--22510, 2023.

\bibitem[Saharia et~al.(2022)Saharia, Chan, Saxena, Li, Whang, Denton,
  Ghasemipour, Ayan, Mahdavi, Lopes, Salimans, Ho, Fleet, and
  Norouzi]{saharia2022imagen}
Chitwan Saharia, William Chan, Saurabh Saxena, Lala Li, Jay Whang, Emily~L.
  Denton, Seyed Kamyar~Seyed Ghasemipour, Burcu~Karagol Ayan, S.~Sara Mahdavi,
  Raphael~Gontijo Lopes, Tim Salimans, Jonathan Ho, David~J. Fleet, and
  Mohammad Norouzi.
\newblock Photorealistic text-to-image diffusion models with deep language
  understanding.
\newblock In \emph{Advances in Neural Information Processing Systems},
  volume~35, 2022.

\bibitem[Simsar et~al.(2025)Simsar, Hofmann, Tombari, and
  Yanardag]{simsar2025loraclr}
Enis Simsar, Thomas Hofmann, Federico Tombari, and Pinar Yanardag.
\newblock {LoRACLR}: Contrastive adaptation for customization of diffusion
  models.
\newblock In \emph{Proceedings of the IEEE/CVF Conference on Computer Vision
  and Pattern Recognition (CVPR)}, pp.\  13189--13198, 2025.

\bibitem[Soboleva et~al.(2026)Soboleva, Alanov, Kuznetsov, and
  Sobolev]{soboleva2026tlora}
Vera Soboleva, Aibek Alanov, Andrey Kuznetsov, and Konstantin Sobolev.
\newblock {T-LoRA}: Single image diffusion model customization without
  overfitting.
\newblock In \emph{Proceedings of the AAAI Conference on Artificial
  Intelligence}, volume~40, pp.\  9051--9059, 2026.
\newblock \doi{10.1609/aaai.v40i11.37861}.

\bibitem[Song et~al.(2021)Song, Sohl-Dickstein, Kingma, Kumar, Ermon, and
  Poole]{song2021sde}
Yang Song, Jascha Sohl-Dickstein, Diederik~P. Kingma, Abhishek Kumar, Stefano
  Ermon, and Ben Poole.
\newblock Score-based generative modeling through stochastic differential
  equations.
\newblock In \emph{International Conference on Learning Representations
  (ICLR)}, 2021.

\bibitem[Struppek et~al.(2023)Struppek, Hintersdorf, and
  Kersting]{struppek2023rickrolling}
Lukas Struppek, Dominik Hintersdorf, and Kristian Kersting.
\newblock Rickrolling the artist: Injecting backdoors into text encoders for
  text-to-image synthesis.
\newblock In \emph{Proceedings of the IEEE/CVF International Conference on
  Computer Vision (ICCV)}, pp.\  4584--4596, 2023.

\bibitem[Vice et~al.(2024)Vice, Akhtar, Hartley, and Mian]{vice2024bagm}
Jordan Vice, Naveed Akhtar, Richard Hartley, and Ajmal Mian.
\newblock {BAGM}: A backdoor attack for manipulating text-to-image generative
  models.
\newblock \emph{IEEE Transactions on Information Forensics and Security},
  19:\penalty0 4865--4880, 2024.
\newblock \doi{10.1109/TIFS.2024.3386058}.

\bibitem[Wang et~al.(2024)Wang, Guo, He, Chen, Zhang, Zhang, and
  Xiang]{wang2024eviledit}
Hao Wang, Shangwei Guo, Jialing He, Kangjie Chen, Shudong Zhang, Tianwei Zhang,
  and Tao Xiang.
\newblock {EvilEdit}: Backdooring text-to-image diffusion models in one second.
\newblock In \emph{Proceedings of the 32nd ACM International Conference on
  Multimedia}, pp.\  3657--3665, 2024.
\newblock \doi{10.1145/3664647.3680689}.

\bibitem[Yang \& Wang(2023)Yang and Wang]{yang2023representation}
Xingyi Yang and Xinchao Wang.
\newblock Diffusion model as representation learner.
\newblock In \emph{Proceedings of the IEEE/CVF International Conference on
  Computer Vision (ICCV)}, pp.\  18938--18949, 2023.

\bibitem[Ye et~al.(2023)Ye, Zhang, Liu, Han, and Yang]{ye2023ipadapter}
Hu~Ye, Jun Zhang, Sibo Liu, Xiao Han, and Wei Yang.
\newblock {IP-Adapter}: Text compatible image prompt adapter for text-to-image
  diffusion models.
\newblock \emph{arXiv preprint arXiv:2308.06721}, 2023.
\newblock \doi{10.48550/arXiv.2308.06721}.

\bibitem[Zhai et~al.(2023)Zhai, Dong, Shen, Pu, Fang, and Su]{zhai2023badt2i}
Shengfang Zhai, Yinpeng Dong, Qingni Shen, Shi Pu, Yuejian Fang, and Hang Su.
\newblock Text-to-image diffusion models can be easily backdoored through
  multimodal data poisoning.
\newblock In \emph{Proceedings of the 31st ACM International Conference on
  Multimedia}, pp.\  1577--1587, 2023.
\newblock \doi{10.1145/3581783.3612108}.

\bibitem[Zhang et~al.(2023)Zhang, Chen, Bukharin, He, Cheng, Chen, and
  Zhao]{zhang2023adalora}
Qingru Zhang, Minshuo Chen, Alexander Bukharin, Pengcheng He, Yu~Cheng, Weizhu
  Chen, and Tuo Zhao.
\newblock Adaptive budget allocation for parameter-efficient fine-tuning.
\newblock In \emph{International Conference on Learning Representations
  (ICLR)}, 2023.

\bibitem[Zhang et~al.(2018)Zhang, Isola, Efros, Shechtman, and
  Wang]{zhang2018lpips}
Richard Zhang, Phillip Isola, Alexei~A. Efros, Eli Shechtman, and Oliver Wang.
\newblock The unreasonable effectiveness of deep features as a perceptual
  metric.
\newblock In \emph{Proceedings of the IEEE Conference on Computer Vision and
  Pattern Recognition (CVPR)}, pp.\  586--595, 2018.
\newblock URL
  \url{https://openaccess.thecvf.com/content_cvpr_2018/html/Zhang_The_Unreasonable_Effectiveness_CVPR_2018_paper.html}.

\bibitem[Zhong et~al.(2024)Zhong, Shen, Wang, Lu, Jiao, Ouyang, Yu, Han, and
  Chen]{zhong2024multilora}
Ming Zhong, Yelong Shen, Shuohang Wang, Yadong Lu, Yizhu Jiao, Siru Ouyang,
  Donghan Yu, Jiawei Han, and Weizhu Chen.
\newblock {Multi-LoRA} composition for image generation.
\newblock \emph{Transactions on Machine Learning Research}, 2024.

\end{thebibliography}
\bibliographystyle{iclr2027_conference}

\appendix
\section{Interaction Derivations}
\label{app:interaction_derivations}

For a generated output mapped by $\Psi$ to the common action space, the behavior attributable specifically to composition is
\begin{equation}
\begin{split}
I_{ij}^{\mathrm{out}}(x,\xi)={}&y_{S_i+P_j}(x,\xi)-y_{S_i}(x,\xi)\\
&-y_{P_j}(x,\xi)+y_{\varnothing}(x,\xi),
\qquad y_A=\Psi(F_A(x;\xi)).
\label{eq:output_interaction}
\end{split}
\end{equation}
Using the same prompt and seed, the corresponding feature interaction at layer $\ell$ and timestep $t$ is
\begin{equation}
I_{ij}^{(\ell,t)}=
h_{S_i+P_j}^{(\ell,t)}-h_{S_i}^{(\ell,t)}-h_{P_j}^{(\ell,t)}
+h_{\varnothing}^{(\ell,t)}.
\label{eq:feature_interaction}
\end{equation}
A purely additive composition makes these differences vanish. Training instead aligns diagonal interactions with the pair action and suppresses off-diagonal interactions.

The stable scalar interface follows from the affine LayerNorm preceding the multiplicative block. For $z_t^{(\ell)}=\gamma^{(\ell)}\odot\widehat h_t^{(\ell)}+\beta^{(\ell)}$ and $\mathbf{1}^{\top}\widehat h_t^{(\ell)}=0$, choosing $v^{(\ell)}=\kappa{\gamma^{(\ell)}}^{-1}$ gives
\begin{equation}
b_t^{(\ell)}={v^{(\ell)}}^\top z_t^{(\ell)}
=\kappa\mathbf{1}^\top\widehat h_t^{(\ell)}
+{v^{(\ell)}}^\top\beta^{(\ell)}=b_0^{(\ell)}.
\label{eq:constant_interface}
\end{equation}
This bias-free projection need only keep ordinary variation within the cancellation region; unstable carrier coordinates are rejected during calibration.

Finally, if $s_i$ and $p_j$ are the local Signature and Payload perturbations and $\Phi$ is the native block transformation, their nonlinear cross-effect is
\begin{equation}
C_{ij}(z)=\Phi(z+s_i+p_j)-\Phi(z+s_i)-\Phi(z+p_j)+\Phi(z).
\label{eq:native_cross_effect}
\end{equation}
It vanishes for a linear block and can be nonzero in GEGLU because the Signature gate displacement interacts multiplicatively with the Payload value/output path. Equation~\ref{eq:released_residual} summarizes a local path under a shared-value approximation, absorbing the value feature and its output projection into the effective vector $u_j^{(\ell)}$. That vector need not be constant across inputs or timesteps; the equation is not an exact expression for the full-network output.

\section{Auxiliary Training Objectives}
\label{app:auxiliary_objectives}

This appendix expands the training discussion in Section~\ref{sec:joint_optimization}, distinguishing the implemented refinement losses from schematic auxiliary design objectives. Expectations are over the relevant prompts, seeds, layers, and timesteps.

\paragraph{Implemented refinement losses.}
Action refinement uses normalized mean squared error (NMSE), masked NMSE, and a direction-alignment loss. The first two penalize prediction discrepancies globally and over masked entries, respectively; the last encourages alignment with the target direction. Not all conceptual objectives below are optimized in every stage; in particular, $\lambda_{\mathrm{off}}=0$ for P3/P4. Appendix~\ref{app:reproducibility} reports the stage-specific losses, weights, and schedules. These training losses are distinct from the image-level Single-Adapter MSE used for evaluation.

\paragraph{Conceptual objective decomposition.}
The intended design constraints over prompts, seeds, compositions, and timesteps $\mathcal{T}$ can be summarized as
\begin{equation}
\begin{split}
    \mathcal{L} ={}&
    \lambda_{\mathrm{diag}}\mathcal{L}_{\mathrm{diag}}
    +\lambda_{\mathrm{off}}\mathcal{L}_{\mathrm{off}}
    +\lambda_{\mathrm{single}}\mathcal{L}_{\mathrm{single}}\\
    &+\lambda_{\mathrm{util}}\mathcal{L}_{\mathrm{util}}
    +\lambda_{\mathrm{traj}}\mathcal{L}_{\mathrm{traj}}
    +\lambda_{\mathrm{cancel}}\mathcal{L}_{\mathrm{cancel}}
    +\lambda_{\mathrm{reg}}\mathcal{L}_{\mathrm{reg}}.
    \label{eq:complete_objective}
\end{split}
\end{equation}
Here $\mathcal{L}_{\mathrm{diag}}$ and $\mathcal{L}_{\mathrm{off}}$ describe matched-action alignment and mismatch suppression. The remaining five terms describe standalone inactivity, utility preservation, trajectory consistency, cancellation, and regularization, respectively. This decomposition organizes the design goals rather than specifying a single seven-term loss jointly optimized in every reported run. The schematic formulations below therefore should not be read as five additional losses necessarily enabled during training.

\paragraph{Single-adapter inactivity.}
Let $a_i$ be a nonnegative differentiable surrogate for the programmed action $r_i$. We suppress that action when either adapter is loaded alone:
\begin{equation}
\begin{split}
    \mathcal{L}_{\mathrm{single}}
    =
    \mathbb{E}_{x,\xi}
    \left[
        \sum_{i=1}^{N}
        \Big(
            a_i\!\left(F_{S_i}(x;\xi)\right)
            +
            a_i\!\left(F_{P_i}(x;\xi)\right)
        \Big)
    \right].
    \label{eq:single_objective}
\end{split}
\end{equation}
This term prevents the Signature from directly carrying the action and the Payload from exposing it without the matched codeword.

\paragraph{Benign-utility preservation.}
Each modified adapter is constrained to reproduce the behavior of its benign precursor:
\begin{equation}
\begin{split}
    \mathcal{L}_{\mathrm{util}}
    =
    \mathbb{E}_{x,\xi}
    \Bigg[
        \sum_{i=1}^{N}
        \Big(
            &d_u\!\left(
                F_{S_i}(x;\xi),
                F_{S_i^{\mathrm{ben}}}(x;\xi)
            \right)\\
            +{}&d_u\!\left(
                F_{P_i}(x;\xi),
                F_{P_i^{\mathrm{ben}}}(x;\xi)
            \right)
        \Big)
    \Bigg],
    \label{eq:utility_objective}
\end{split}
\end{equation}
where $d_u$ may combine pixel, perceptual, and representation distances according to the advertised utility of each adapter.

\paragraph{Trajectory consistency.}
We align matched interactions with layer- and timestep-specific training directions over the selected carriers and denoising timesteps:
\begin{equation}
    \mathcal{L}_{\mathrm{traj}}
    =
    \mathbb{E}_{x,\xi}
    \left[
        \sum_{i=1}^{N}
        \sum_{t\in\mathcal{T}}
        \sum_{\ell\in\mathcal{K}}
        d_{\mathrm{traj}}\!\left(
            I_{ii}^{(\ell,t)},
            d_i^{(\ell,t)}
        \right)
    \right].
    \label{eq:trajectory_objective}
\end{equation}
The direction $d_i^{(\ell,t)}$ may be obtained from a teacher, a target residual, or a differentiable semantic objective; it is used only during training.

\paragraph{Differential cancellation.}
For payload $P_j$, let $\pi_{\mathrm{off},j}$ sample the non-matched states $\{P_j\}\cup\{S_i+P_j:i\neq j\}$, and let $h_A$ denote the activation under sampled state $A$. We preserve cancellation between its signal and reference branches in those states:
\begin{equation}
\begin{split}
    \mathcal{L}_{\mathrm{cancel}}
    =
    \sum_{j=1}^{N}
    \mathbb{E}_{x,\xi,A\sim\pi_{\mathrm{off},j}}
    \Bigg[
        \sum_{t\in\mathcal{T}}
        \sum_{\ell\in\mathcal{K}}
        \left\|
            R_{P_j,+}^{(\ell,t)}(h_A)
            +
            R_{P_j,-}^{(\ell,t)}(h_A)
        \right\|_2^2
    \Bigg].
    \label{eq:cancellation_objective}
\end{split}
\end{equation}
This objective does not constrain the matched state, where the Signature is expected to break cancellation and release the action.

\paragraph{Regularization.}
For clarity, we decompose the regularizer as
\begin{equation}
    \mathcal{L}_{\mathrm{reg}}
    =\mu_{\theta}\mathcal{R}_{\theta}
    +\mu_{\alpha}\mathcal{R}_{\alpha}
    +\mu_{h}\mathcal{R}_{\mathrm{drift}}
    +\mu_{R}\mathcal{R}_{\mathrm{energy}},
    \label{eq:regularization_objective}
\end{equation}
where the four terms represent penalties on implanted LoRA weights, Signature amplitude, ordinary-state activation drift, and carrier residual energy, respectively. This decomposition describes possible regularization roles; it does not specify four separately enabled penalties in the reported runs.

\section{Reproducibility Details}
\label{app:reproducibility}

\paragraph{Modules and ranks.}
The benign LoRAs target \texttt{to\_q}, \texttt{to\_k}, \texttt{to\_v}, and
\texttt{to\_out.0}; the last name denotes the linear attention-output projection in Diffusers.
Carrier updates target \texttt{ff.net.0.proj}, the joint GEGLU gate/value input projection, and
\texttt{ff.net.2}, its output projection. Payload projections use rank 4. The Signature uses
rank 5 at \texttt{ff.net.0.proj} for an additional constant-gate dimension and rank 4 at
\texttt{ff.net.2}; the P1 \texttt{mid\_t7} carrier uses ranks 4/3 because it has three lanes.

\paragraph{Schedules.}
Benign LoRAs are trained for 1,800 AdamW steps at $5\times10^{-5}$ and refined for 1,200 steps at $2\times10^{-5}$. P1/P2 use a staged warm start totaling 10,100 updates; their final 2,500-step joint refinement uses $3\times10^{-4}$ for the action and $2\times10^{-4}$ for the reader. P3 uses 3,000 initial action steps at $1.5\times10^{-3}$ and 1,400 refinement steps at $10^{-3}$; P4 uses 1,400 refinement steps at $10^{-3}$. Action refinement minimizes NMSE plus $0.5$ masked NMSE and $0.5$ direction loss. AdamW uses $(\beta_1,\beta_2)=(0.9,0.99)$, weight decay $10^{-4}$, and gradient clipping of 1.0 for utility parameters and 2.0 for action/reader parameters; P3/P4 use $\lambda_{\mathrm{off}}=0$.

\paragraph{Export coefficients.}
For P1--P4, respectively, gate coefficients are $\{29,35,29,46\}$, reader coefficients are $\{0.08,0.06,0.08,0.05\}$, action coefficients are $\{0.33,0.45,0.48,0.462\}$, and value-cancellation coefficients are $\{0.85,0.90,0.85,0.85\}$.

\section{Per-Cell Pairing Metrics}
\label{app:pairing_metrics}

This appendix expands the pair-selectivity results in Figure~\ref{fig:pairing_matrix}. Each cell $S_i+P_j$ is evaluated on the same $n=100$ prompt--seed cases. Diagonal cells ($i=j$) are the intended matched pairs; off-diagonal cells ($i\neq j$) are mismatched pairs that should remain inactive.

\paragraph{Standalone fidelity.}
For each Signature or Payload, CLIP distance measures semantic deviation between the implanted adapter and its clean precursor, while pixel MSE measures their image-level deviation under the same prompt and seed. Lower values indicate better preservation of the adapter's advertised standalone behavior. Because these quantities depend only on the individual adapter, Table~\ref{tab:appendix_standalone} places each intended Signature--Payload pair on one row so that the two adapters can be compared directly without repeating their standalone values across mismatched compositions.

\paragraph{Pair-conditioned behavior.}
Table~\ref{tab:appendix_pair_metrics} reports Target Shift and off-pair activation rate for every Signature--Payload composition under the $N=4$ and $N=8$ configurations. Target Shift measures movement of the composed output toward the designated target. A large positive diagonal value indicates effective activation by the intended pair, whereas an off-diagonal value near zero indicates that a mismatched pair remains suppressed. The \emph{off-pair activation rate} is the percentage of the 100 cases in a mismatched cell that nevertheless activate the target---that is, the empirical probability of an unintended pairing success. Lower is better. It is undefined for diagonal cells, which are intended to activate and are therefore marked by an em dash. The aggregate rate pools all off-diagonal cases: 1,200 cases for $N=4$ and 5,600 cases for $N=8$.

\begin{table}[htbp]
\caption{Standalone fidelity of the intended Signature--Payload pairs used in Figure~\ref{fig:pairing_matrix}. Each row compares the two members of one matched pair with their respective clean precursors; CLIP distance and pixel MSE are both lower-is-better.}
\label{tab:appendix_standalone}
\centering
\small
\setlength{\tabcolsep}{8pt}
\begin{tabular}{lcccc}
\toprule
& \multicolumn{2}{c}{Signature-only} & \multicolumn{2}{c}{Payload-only} \\
& \multicolumn{2}{c}{$(S_i^{\mathrm{mod}}, S_i^{\mathrm{clean}})$}
& \multicolumn{2}{c}{$(P_i^{\mathrm{mod}}, P_i^{\mathrm{clean}})$} \\
\cmidrule(lr){2-3}\cmidrule(lr){4-5}
Pair ID $i$ & CLIP dist. $\downarrow$ & Pixel MSE $\downarrow$
& CLIP dist. $\downarrow$ & Pixel MSE $\downarrow$ \\
\midrule
\multicolumn{5}{c}{\textbf{Pair-set size }$\mathbf{N=4}$} \\
\cmidrule(lr){1-5}
1 & 0.013 & 0.006 & 0.011 & 0.003 \\
2 & 0.012 & 0.005 & 0.008 & 0.002 \\
3 & 0.013 & 0.005 & 0.008 & 0.002 \\
4 & 0.011 & 0.005 & 0.008 & 0.002 \\
\midrule
\multicolumn{5}{c}{\textbf{Pair-set size }$\mathbf{N=8}$} \\
\cmidrule(lr){1-5}
1 & 0.017 & 0.011 & 0.015 & 0.005 \\
2 & 0.019 & 0.012 & 0.019 & 0.006 \\
3 & 0.019 & 0.011 & 0.016 & 0.005 \\
4 & 0.015 & 0.010 & 0.014 & 0.005 \\
5 & 0.017 & 0.011 & 0.012 & 0.004 \\
6 & 0.015 & 0.010 & 0.013 & 0.004 \\
7 & 0.018 & 0.011 & 0.013 & 0.005 \\
8 & 0.015 & 0.009 & 0.010 & 0.004 \\
\bottomrule
\end{tabular}
\par\vspace{4pt}
\begin{minipage}{\linewidth}
\tiny\raggedright
\textit{Note.} The Signature columns report CLIP distance and pixel MSE between outputs generated with $S_i^{\mathrm{mod}}$ alone and with $S_i^{\mathrm{clean}}$ alone. The Payload columns report the same metrics for $P_i^{\mathrm{mod}}$ alone versus $P_i^{\mathrm{clean}}$ alone, using the same prompt and seed. Each row groups the two members of an intended pair; these are standalone-adapter measurements, not measurements of their co-loaded output.
\end{minipage}
\end{table}

\begin{table}[p]
\caption{Pair-conditioned metrics for every Signature--Payload composition. For each Signature, the first row gives Target Shift and the second gives off-pair activation rate. Diagonal entries are intended matched pairs, so their off-pair rate is not applicable and is marked by an em dash.}
\label{tab:appendix_pair_metrics}
\centering
\footnotesize
\setlength{\tabcolsep}{1.3pt}
\begin{tabular*}{\linewidth}{@{\extracolsep{\fill}}llrrrrrrrr@{}}
\toprule
\multicolumn{10}{c}{\textbf{Pair-set size }$\mathbf{N=4}$} \\
\cmidrule(lr){1-10}
Signature & Metric & \multicolumn{2}{c}{$P_1$} & \multicolumn{2}{c}{$P_2$} & \multicolumn{2}{c}{$P_3$} & \multicolumn{2}{c}{$P_4$} \\
\midrule
\multirow{2}{*}{$S_1$} & Shift & \multicolumn{2}{c}{0.085} & \multicolumn{2}{c}{0.000} & \multicolumn{2}{c}{$-0.001$} & \multicolumn{2}{c}{0.000} \\
& Off (\%) & \multicolumn{2}{c}{---} & \multicolumn{2}{c}{1.0} & \multicolumn{2}{c}{0.0} & \multicolumn{2}{c}{0.0} \\
\addlinespace[1pt]
\multirow{2}{*}{$S_2$} & Shift & \multicolumn{2}{c}{0.001} & \multicolumn{2}{c}{0.037} & \multicolumn{2}{c}{$-0.001$} & \multicolumn{2}{c}{0.000} \\
& Off (\%) & \multicolumn{2}{c}{3.0} & \multicolumn{2}{c}{---} & \multicolumn{2}{c}{0.0} & \multicolumn{2}{c}{0.0} \\
\addlinespace[1pt]
\multirow{2}{*}{$S_3$} & Shift & \multicolumn{2}{c}{$0.000$} & \multicolumn{2}{c}{0.000} & \multicolumn{2}{c}{0.061} & \multicolumn{2}{c}{0.000} \\
& Off (\%) & \multicolumn{2}{c}{1.0} & \multicolumn{2}{c}{0.0} & \multicolumn{2}{c}{---} & \multicolumn{2}{c}{0.0} \\
\addlinespace[1pt]
\multirow{2}{*}{$S_4$} & Shift & \multicolumn{2}{c}{0.000} & \multicolumn{2}{c}{$-0.001$} & \multicolumn{2}{c}{$-0.001$} & \multicolumn{2}{c}{0.025} \\
& Off (\%) & \multicolumn{2}{c}{1.0} & \multicolumn{2}{c}{0.0} & \multicolumn{2}{c}{0.0} & \multicolumn{2}{c}{---} \\
\addlinespace[4pt]
\midrule
\multicolumn{10}{c}{\textbf{Pair-set size }$\mathbf{N=8}$} \\
\cmidrule(lr){1-10}
Signature & Metric & $P_1$ & $P_2$ & $P_3$ & $P_4$ & $P_5$ & $P_6$ & $P_7$ & $P_8$ \\
\midrule
\multirow{2}{*}{$S_1$} & Shift & 0.113 & $-0.002$ & $-0.002$ & $-0.003$ & 0.003 & $-0.004$ & $-0.004$ & 0.003 \\
& Off (\%) & --- & 0.0 & 0.0 & 0.0 & 0.0 & 0.0 & 0.0 & 0.0 \\
\addlinespace[0.5pt]
\multirow{2}{*}{$S_2$} & Shift & $0.000$ & 0.098 & 0.000 & $-0.009$ & $0.000$ & $-0.005$ & $-0.004$ & 0.003 \\
& Off (\%) & 0.0 & --- & 0.0 & 0.0 & 0.0 & 0.0 & 0.0 & 0.0 \\
\addlinespace[0.5pt]
\multirow{2}{*}{$S_3$} & Shift & $0.000$ & $-0.005$ & 0.139 & 0.001 & 0.001 & $-0.003$ & $-0.006$ & 0.005 \\
& Off (\%) & 0.0 & 0.0 & --- & 0.0 & 0.0 & 0.0 & 0.0 & 0.0 \\
\addlinespace[0.5pt]
\multirow{2}{*}{$S_4$} & Shift & 0.004 & $0.000$ & 0.001 & 0.108 & 0.004 & $-0.003$ & $-0.007$ & 0.004 \\
& Off (\%) & 0.0 & 0.0 & 0.0 & --- & 0.0 & 0.0 & 0.0 & 0.0 \\
\addlinespace[0.5pt]
\multirow{2}{*}{$S_5$} & Shift & $-0.001$ & $-0.001$ & 0.002 & $-0.001$ & 0.096 & $-0.006$ & $-0.004$ & $-0.001$ \\
& Off (\%) & 0.0 & 0.0 & 0.0 & 0.0 & --- & 0.0 & 0.0 & 0.0 \\
\addlinespace[0.5pt]
\multirow{2}{*}{$S_6$} & Shift & 0.004 & 0.000 & 0.002 & $-0.003$ & 0.003 & 0.123 & $-0.005$ & 0.001 \\
& Off (\%) & 0.0 & 0.0 & 0.0 & 0.0 & 0.0 & --- & 0.0 & 0.0 \\
\addlinespace[0.5pt]
\multirow{2}{*}{$S_7$} & Shift & 0.003 & 0.002 & 0.001 & $-0.005$ & $-0.001$ & $-0.005$ & 0.117 & 0.002 \\
& Off (\%) & 0.0 & 0.0 & 0.0 & 0.0 & 0.0 & 0.0 & --- & 0.0 \\
\addlinespace[0.5pt]
\multirow{2}{*}{$S_8$} & Shift & 0.002 & $-0.002$ & 0.001 & 0.001 & $-0.002$ & $-0.005$ & $-0.006$ & 0.118 \\
& Off (\%) & 0.0 & 0.0 & 0.0 & 0.0 & 0.0 & 0.0 & 0.0 & --- \\
\bottomrule
\end{tabular*}
\end{table}

\end{document}